# Hyperbolic Metamaterials and Coupled Surface Plasmon Polaritons: comparative analysis

TENGFEI LI AND JACOB B. KHURGIN*

*Johns Hopkins University, Department of Electrical and Computer Engineering, 3400 N. Charles St, Baltimore, MD 21218*
*Corresponding author: jakek@jhu.edu*

**We investigate the optical properties of sub-wavelength layered metal/dielectric structures, also known as hyperbolic metamaterials (HMMs), using exact analytical Kronig Penney (KP) model. We show that hyperbolic isofrequency surfaces exist for all combinations of layer permittivities and thicknesses, and the largest Purcell enhancements (PE) of spontaneous radiation are achieved away from the nominally hyperbolic region. Detailed comparison of field distributions, dispersion curves, and Purcell factors (PF) between the HMMs and Surface Plasmon Polaritons (SPPs) guided modes in metal/dielectric waveguides demonstrates that HMMs are nothing but weakly coupled gap or slab SPPs modes. Broadband PE is not specific to the HMMs and can be easily attained in single thin metallic layers. Furthermore, large wavevectors and PE are always combined with high loss, short propagation distances and large impedances; hence PE in HMMs is essentially a direct coupling of the energy into the free electron motion in the metal, or quenching of radiative lifetime. PE in HMMs is not related to the hyperbolicity per se but is simply the consequence of the strong dispersion of permittivity in the metals or polar dielectrics, as our conclusions are relevant also for the infrared HMMs occurring in nature. When it comes to enhancement of radiating processes and field concentrations, HMMs are not superior to far simpler plasmonic structures. © 2016 Optical Society of America**

## 1. INTRODUCTION

Recent years have seen an exponentially growing interest in the artificial material made up from sub-wavelength dielectric and/or metal elements, which are known as optical metamaterials [1-3]. Among many classes of optical metamaterials, the ones that attracted special attention are hyperbolic metamaterials (HMMs), characterized by unusual properties of their dielectric permittivity tensor whose diagonal components have different signs [4]. As shown in Fig. 1, HMMs can comprise either alternating metal and dielectric layers or an array of metal nanowires embedded in a dielectric. In addition to the synthetic HMMs, the naturally occurring hyperbolic materials (HMs) are also known to exist [5]. Fig. 1 (b) shows the isofrequency surfaces (IFS) of normal dielectric, two types of HMMs and the elliptical IFS of the conventional anisotropic dielectric. The salient feature of hyperbolic IFS is that the wavevectors in them span the unlimited range which portends their potential use in the fields of high resolution imaging [6, 7], photonic density of states manipulation [8], negative refraction [9–11], spontaneous emission engineering [12–15], epsilon-near-zero metamaterial [16], and thermal emission engineering [17, 18]. There have been a number of remarkable experimental results with HMMs, including high resolution imaging using so-called "hyperlens "[6, 7], and numerous works on enhancement of spontaneous emission via Purcell effect inherent to the materials supporting large wavevector [19–21]. The latter results, however, have been far from being superior to the PE achieved with other metal/dielectric nanostructures, such as optical nano-antennas of all kinds [22–25]. While it is clear that the difference between predicted and experimentally observed performance is most probably related to high loss in the metal, a detailed theoretical study of HMMs would go a long way towards answering the question of whether they do have any tangible advantage over the more explored plasmonic structures. When it comes to the theoretical analysis of HMMs to date, the well-tried effective medium theory (EMT) has been relied upon the most [21, 26–28], as well as more complex transfer matrix [29–31] and Green's function methods [28, 32], which predict Purcell factors (PF) reaching values exceeding $10^5$ [28], conversely, as mentioned above, experimentally reported values do not exceed 80 [19, 20, 33, 34]. Numerical FDTD simulation [34, 35] has also been applied to HMMs, showing, once again, the predicted PE far exceeding the observed values. Besides being computationally cumbersome, neither one of these aforementioned "beyond the EMT" methods directly produce IFS in k-space, while FDTD method fails to offer any physical insight into the picture. Based on the analysis of the rich body of theoretical HMMs work, there are still several questions waiting to be answered: (1) According to the EMT, the hyperbolic IFS exists only for certain relations of layer permittivities and thicknesses. While this prediction must be correct in the limit of infinitely thin layers, it has already been shown in [31] that the elliptical and hyperbolic dispersion region can overlap; (2) While the giant enhancement of emission rates into the HMMs has been predicted, the enhancement of the rate of radiation coming out of HMMs, i.e., external rather than internal efficiency has not been thoroughly investigated; (3) Spatial dependence of Purcell effect, critical from practical point of view, has not been given proper attention; (4) Most importantly, as any plasmonic (metal/dielectric) structure exhibits the same features as HMMs, namely strong field confinement and the ability to support large wavevectors and PE, it is crucial to define the connection between HMMs and the more conventional slab and gap SPPs. In order to address these questions, we shall use an analytical model that combines simplicity with precision letting us investigate all the relevant properties of HMMs in great detail and comparing them with SPPs. While the results of our work can be used in evaluation of HMMs in any potential application, we mainly focus on the enhancement of the spontaneous emission. Furthermore, while our focus is on the man-made HMMs comprising metal/dielectric layers, in the end we show that many of our conclusions are also held for natural HMs [5].

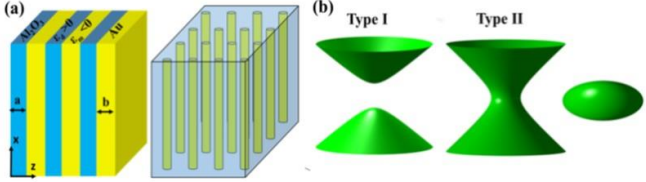

Fig. 1. (a) Two configurations of HMMs; (b) IFS for type I and type II HMMs and normal dielectric medium.

## 2. KROGIG-PENNEY MODEL OF HMMS

The Kronig-Penney (KP) model was developed in the 1930's [36] in order to provide a simple explanation of the formation of the band structure in the periodic lattice. Obviously, any attempt to approximate the real crystal potential by a periodic sequence of one-dimensional wells and barriers is bound to lack precision. Thus the KP model has been largely relegated to condensed matter textbooks, until it had enjoyed a brief renaissance in 1980's when it was successfully applied to semiconductor superlattices [37, 38], where the KP potential bears a much closer resemblance to the actual superlattice potential. While the KP model is always an approximation in condensed matter physics, in periodic photonic structures the KP model is an exact one. Changes in the dielectric permittivity are perfectly well described by the square wave function. The KP has been successfully used in one-dimensional photonic crystals [39] and it appears to be a good choice for the study of HMMs. The KP model is not computationally heavy and readily provides physical characteristics, such as, IFS, dispersion curves and field shapes. Given these attractive features of the KP model, it appears to be perfectly suited to our task of providing insight into the physics of HMMs.

As an example of a type I HMMs, we consider the structure in Fig. 1(a), consisting of the $Al_2O_3$ [7] layers with thickness $a$ and constant permittivity in the visible range $\varepsilon_d = 3.61$ alternating with the silver layers of thickness $b$. The dielectric constant of Ag in the visible range can be fitted into the Drude model: $\varepsilon_m(\omega) = 1 - \omega_p^2 / (\omega^2 + i\omega\gamma_m(\omega))$, where $\omega_p = 1.36 \times 10^{16} S^{-1}$ [40] is the bulk plasma frequency and the scattering rate, $\gamma_m$, is frequency dependent. According to [40], $\gamma_m = 8.475 \times 10^{13} S^{-1}$ at $\lambda = 500 nm$. In order to facilitate calculations and also make the physics more transparent, all the distances and wavevectors are normalized to the wavevector in the dielectric, $k_d = \varepsilon_d^{1/2}\omega/c$, as $x' = xk_d$, so the spatial derivative becomes $\nabla' = k_d^{-1}\nabla$. In order to express the magnetic field in the same units (V/m) as the electric field, the magnetic field is normalized to the impedance in the dielectric, $\eta_d = \sqrt{\mu_0/(\varepsilon_0\varepsilon_d)}$, as $H_y' = H_y\eta_d$. This normalization allows us to define relative (to the dielectric) local impedance as $\eta'(z) = E(z)/H'(z)$. As shown further on, relative impedance defines most of the of the characteristics of the modes in HMMs, such as, the degree of confinement, velocity of propagation,

Purcell enhancement and propagation distance. (see Section 1 of Supplement 1).

In the KP model [36], the normalized magnetic field of TM mode in HMMs can be expressed as:

$$H_y' = \begin{cases} \left(Ae^{K'z'} + Be^{-K'z'}\right)e^{ik_x'x'}; & 0 < z' < a' \\ \left(Ce^{Q'z'} + De^{-Q'z'}\right)e^{ik_x'x'}; & -b' < z' < 0 \end{cases} \quad (1)$$

where $k_x'$ is the normalized lateral wavevector, and $K'$, $Q'$ are the normalized decay constants in the dielectric and metal respectively, are related as:

$$Q'^2 - K'^2 = 1 + |\varepsilon_m'|, \quad (2)$$

where $\varepsilon_m' = \varepsilon_m / \varepsilon_d$ is the relative (to dielectric) metal dielectric constant. The amplitudes $A$ through $D$ are defined by applying the periodic boundary condition $H_y(z'+a'+b') = H_y(z')e^{[ik_z'(a+b)]}$ to Eq. 1 and Eq. 3 for the lateral electric field derived from Eq. 1.

$$E_x = \begin{cases} -iK'\left(Ae^{K'z'} - Be^{-K'z'}\right)e^{ik_x'x'}; & 0 < z' < a' \\ -\dfrac{iQ'}{|\varepsilon_m'|}\left(Ce^{Q'z'} - De^{-Q'z'}\right)e^{ik_x'x'}; & -b' < z' < 0 \end{cases} \quad (3)$$

With these boundary conditions, one obtains the characteristic equation of the propagating mode in the HMMs (see the Section 2 of Supplement),

$$\cos\left(k_z'(a'+b')\right) = -\frac{1}{2}\left(\frac{Q'}{K'|\varepsilon_m'|} + \frac{K'|\varepsilon_m'|}{Q'}\right)\sinh(K'a')\sinh(Q'b') + \cosh(K'a')\cosh(Q'b') \quad (4)$$

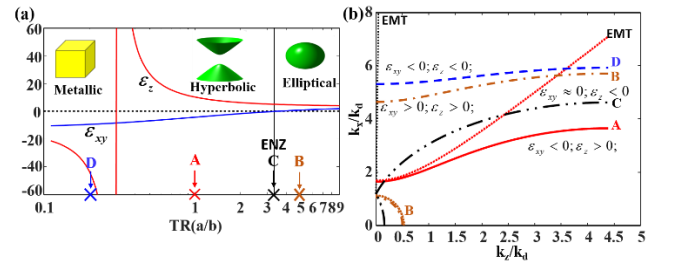

Fig. 2. (a) Dielectric constants for different $TR$ from EMT; (b) IFS for different TR by using Kronig-Penney model.

which can also be obtained by the transition matrix method [31]. Each solution of Eq. 4 yields the value of transverse wavevector $k_z'$ for each lateral wave-vector $k_x'$. Thus generating the points on the IFS, as shown in Fig. 2(b) for the wavelength of $\lambda = 500 nm$ (this wavelength is 67% longer that the wavelength of SPPs $\lambda_{SPPs} \approx 300 nm$ on the Ag-$Al_2O_3$ interface), for four different values of the thickness ratio $TR = a/b$, corresponding to four different classes of the effective medium. As shown in Fig. 2(a), large $TR$ indicates that both lateral and transverse effective dielectric constants are positive and the effective medium is conventional (elliptical) dielectric. As $TR$ decreases, first, the lateral dielectric constant $\varepsilon_{xy}$ changes sign indicating that the medium becomes hyperbolic, and then the transverse dielectric constant $\varepsilon_z$ also becomes negative indicating that the material effectively reaches metallic stage. Let us now consider the IFS graphs corresponding to these regions.

For case A, we consider $TR = 1$ for which EMT predicts $\varepsilon_{xy} = -4.20, \varepsilon_z = 10.32$, placing the structure squarely in the hyperbolic region with IFS shown in Fig. 2(b), following the red dotted line A. The IFS calculated using the KP method (solid red line A) shows hyperbolic dispersion, but we can see that the two results match each other only when k vector is rather small, not exceeding 10% of the Brillouin zone (BZ), after which the slope of the KP line gradually decreases to zero as a consequence of the reflection at the edge of BZ.

For case B, we consider $TR = 5$ for which effective permittivities are both positive, $\varepsilon_{xy} = 1, \varepsilon_z = 4.61$, hence EMT predicts that there should be only elliptical IFS, as shown by the brown dotted line B in Fig. 2(b). However, the KP solution (brown dash-dot line) demonstrates that the hyperbolic and elliptical IFSs co-exist, and the elliptical part of the KP model matches the EMT very well.

Next, we consider the borderline case C with $TR = 3.33$ and the effective permittivities $\varepsilon_{xy} = -7.9 \times 10^{-4}, \varepsilon_z = 5.20$, which agrees with the definition of the epsilon-near-zero (ENZ) metamaterials which have been studied extensively [41–45]. According to EMT the transverse wavevector $k_z$ can become arbitrarily small, indicating the constant phase extending along the $z$ axis. This observation is confirmed by the EMT IFS rendered by the nearly vertical dotted black line in Fig. 2(b). However, the KP model shown as a black dashed line which takes into account granularity results in the IFS is quite different from the EMT predictions. The IFS has both hyperbolic and elliptical regions which nearly touch each other and, obviously, all the transverse wavevectors within BZ are allowed, as one would be expecting from Floquet-Bloch theorem [46]. Clearly, to achieve "true" ENZ, one must either use much thinner layers or revert to bulk highly doped materials with tunable plasma frequency such as AlZnO [47].

Finally, we consider case D with $TR = 0.2$, comprising very thin dielectric "gaps" sandwiched between thick metal layers. Unsurprisingly EMT predicts two negative effective permittivities $\varepsilon_{xy} = -9.41, \varepsilon_z = -43.11$, making the HMMs an effective metal that cannot support propagating waves in any direction. According to KP method, and in full agreement with Floquet-Bloch theorem, the hyperbolic-like solutions still exist with IFS becoming more and more horizontal as TR further decreases. In fact, the IFS for the nominally elliptic case B and for nominally metallic case D look very similar to each other. They both move "higher" and become more "horizontal" as $TR$ goes to either 0 or infinity. This means that for either very large or very small $TR$, we are dealing with the waves that propagate mostly in only the lateral direction. This corresponds to the behavior of "weakly coupled" modes in the arrays of dielectric waveguides [48]. Clearly, the modes of HMMs can be thought of as coupled modes of plasmonic waveguides. In the metallic region D of small $TR$ those modes are the coupled gap SPPs modes. In the elliptic region B with large $TR$, they are coupled slab SPPs modes [49]. To further investigate this analogy, we must first explore HMMs characteristics beyond IFS curves.

## 3. FIELDS, ENERGY DENSITY AND POYNTING VECTOR

Let us now look at the spatial distribution of the field components as well as the energies and Poynting vectors. The electric and magnetic energy densities (see Section 3 of Supplement 1) are $U_E(z) = \frac{1}{4}\varepsilon_0 \varepsilon_d \varepsilon_g' |E|^2$ and $U_H(z) = \frac{1}{4}\varepsilon_0 \varepsilon_d |H_y'|^2$, where $\varepsilon_g' = \varepsilon_d^{-1} \partial(\omega \varepsilon_{m(d)})/\partial\omega$ is the normalized "group" permittivity (see Section 2 of Supplement 1). Then the total energy density, normalized to the energy of plane wave in the dielectric can be written as $U_T = \frac{1}{2}(\varepsilon_g'(z) + \eta'^{-2})$. For the plane wave propagating in the unconstrained lossless dielectric $\varepsilon_g' = \eta' = 1$ and $U_H'(z) = U_E'(z)$, hence $\eta'$ tells us important facts about the energy balance in HMMs.

According to [50], when magnetic energy is much less than electric one, significant portion of energy gets stored in the kinetic motion of free electrons in the metal, which leads to ohmic loss. Therefore, large $\eta'$ is invariably associated with large loss.

The normalized (to the plane wave in the dielectric) Poynting vector components are calculated as $S_{x(z)}' = \text{Re}(E_{z(x)} H_y^*)/|E|^2 \sim e_{xz}/\eta'(z)$, where $e_{xz} = E_{x(z)}/E$ is the projection of the unit vector $\vec{e}$ indicating the field polarization. With that, one can easily get the normalized (to velocity in the dielectric $v_d = c/\varepsilon_d^{1/2}$) energy velocity, $v_{ex(z)}'(z) = S_{x(z)}'/U_T'(z) = 2e_{xz}/[\eta'^{-1}(z)(\varepsilon_g' + \eta'^{-2}(z))]$. Polarization and position dependent PF [14] is calculated as an integral over the IFS surface (see the second Section of Supplement 1).

$$PF_{x(z)} = \frac{1}{2}\int \frac{E_{x(z)}^2(z)}{\langle \varepsilon_g' E^2 + H'^2 \rangle} k_x' \frac{1}{v_{gx}'} dk_z' \quad (5)$$

The energy averaging in the denominator is over one period of HMMs and $v_{gx}' = \omega^{-1}\partial\omega/\partial k_x'$ is the lateral group velocity normalized to the speed of light in the dielectric. Note that there are three factors behind PE: field enhancement near the metal/dielectric interface, large wavevector, and, most importantly, low group velocity. Let us now explore the HMMs with $TR = 1$, i.e., made up by metal and dielectric layers of equal thickness with the IFS shown in Fig. 3(a). In Fig. 3(b) we show the spatial dependence of the PF which reaches 25 for this configuration. Then we choose two points on the IFS for our investigation.

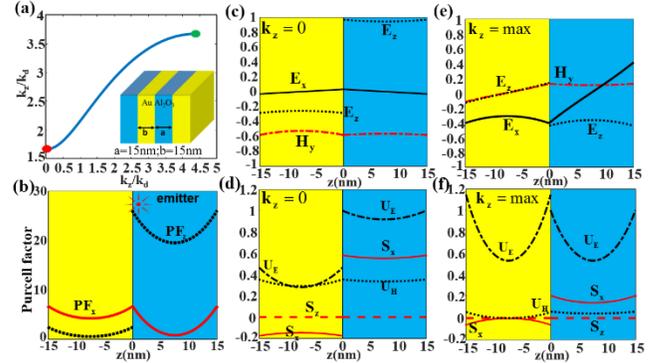

Fig. 3. (a) The IFS at l = 500nm for $TR = 1$; (b) PF as a function of the position of emitter in the dielectric for two polarizations; (c) Fields and (d) Energies and Poynting vector for the minimum value of transverse wavevector; (e, f) The same for the maximum value of transverse wavevector.

For the first point with transverse wavevector $k_z = 0$, the fields and energies are plotted in Fig. 3(c) and (d). One can see that magnetic field is symmetric inside the metal (similar to the so-called "long range SPPs mode in slab waveguide [51, 52]), that the energy is mostly contained inside the dielectric and the magnetic energy is about a factor of two less than electric energy. Also, one can see that the energy propagates in opposite directions in the metal and dielectric. As we move to high transverse wavevectors near the edge of BZ (Fig. 3(e, f)), the symmetry of the mode changes as magnetic field becomes anti-symmetric inside the metal. It is known that in slab waveguides this mode, often referred to as "short range plasmon" [49], can extend to large lateral wavevectors but also suffers from the large loss. Indeed, one can see from Fig. 3(f) that nearly 50% of the energy is contained inside the metal hence the loss is expected to be high [50]. Furthermore, notice that $U_H \ll U_E$ which indicates that the energy balance in the mode is now maintained in a different way as the energy oscillates between the "capacitance" of dielectric and "kinetic inductance" of moving free electrons in the metal. Naturally, the moving electrons always dissipate energy and the high loss ensues. Finally, notice that in the virtual absence of magnetic field, the Poynting vector is small and so should be the group velocity. Then according to Eq. 5, the modes with large $k$ are expected to contribute disproportionally to the density of states and PE. However, they also are expected to have higher loss and shorter propagation distances.

## 4. EFFECTIVE PARAMETERS

To further explore the consequences of the observations made above, we define a number of k-vector dependent parameters, such as the effective mode loss, $\gamma_{eff}(k_z') = f_m(k_z')\gamma_m(k_z')$, where $f_m = \int_{-b'}^{0} U_T' dz' / \int_{-b'}^{a'} U_T' dz'$ is the energy fraction in metal; the effective impedance $\eta_{eff} = \int_{-b'}^{a'} |E(z)| dz' / \int_{-b'}^{a'} |H(z)| dz'$, two components of normalized (to the speed of light in the dielectric) energy velocity $v_{ex(z)}' = \int_{-b'}^{a'} S_{x(z)}' dz' / \int_{-b'}^{a'} U_T' dz'$, and the propagation length calculated by dividing the energy velocity by the effective loss $L_{x(z)} = v_{ex(z)}' / \gamma_{eff}$. (see the fifth Section of Supplement 1). In addition, we introduce the differential PF (simply the integrand in Eq. 5)

$$PF_{x(z)}'(k_z') = \frac{1}{2} \frac{E_{x(z)}^2(z)}{\langle \varepsilon_g' E^2 + H'^2 \rangle} k_x' \frac{1}{v_{gx}'}, \qquad (6)$$

Which describes the relative contribution of the states, with a given transverse wavevector, to the density of states and PE.

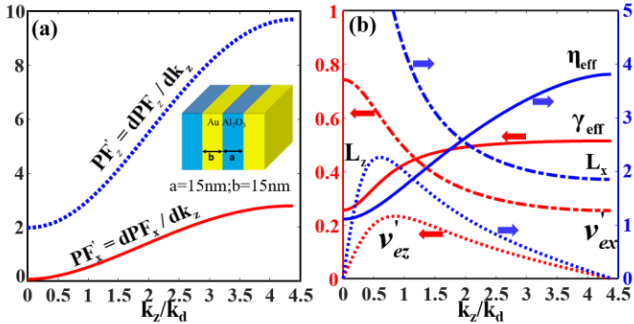

Fig. 4. Change of differential PF (a), and effective parameters (b), with wavevector at *TR* = 1.

From Fig. 4(a) we can see that the differential PF increases with the increase of wavevector, especially when the emitting dipole is polarized along z axis (which is no wonder given TM character of the waves in HMMs). At the same time from Fig. 4(b) we can see that the effective loss and impedance also increase, while the effective energy velocity and propagation length decrease. For example, at large wavevectors the propagation length is only two wavelengths in the dielectric, i.e. about 530nm. Hence most of the "additional" radiation caused by PE actually couples into the lossy modes that do not propagate far, and, moreover, get reflected at HMMs surface due to their large effective impedance.

## 5. MEAN PARAMETERS

Next we explore the properties of layered materials throughout all three regimes (metallic, hyperbolic and dielectric) defined by the *TR*. To facilitate this study, we define the "mean" parameters by weighing the effective parameters over the PF. The mean loss, $\langle \gamma_{eff} \rangle = \int \gamma_{eff} PF_z' dk_z' / \int PF_z' dk_z'$, the mean impedance $\langle \eta_{eff} \rangle = \int \eta_{eff} PF_z' dk_z' / \int PF_z' dk_z'$, the mean energy velocity $\langle v_{ex(z)}' \rangle = \int v_{ex(z)}' PF_z' dk_z' / \int PF_z' dk_z'$, and the mean propagation length $\langle L_{x(z)} \rangle = \int L_{x(z)} PF_z' dk_z' / \int PF_z' dk_z'$, are all the characteristics of the average mode into which the emission takes place (see Section 5 of Supplement 1).

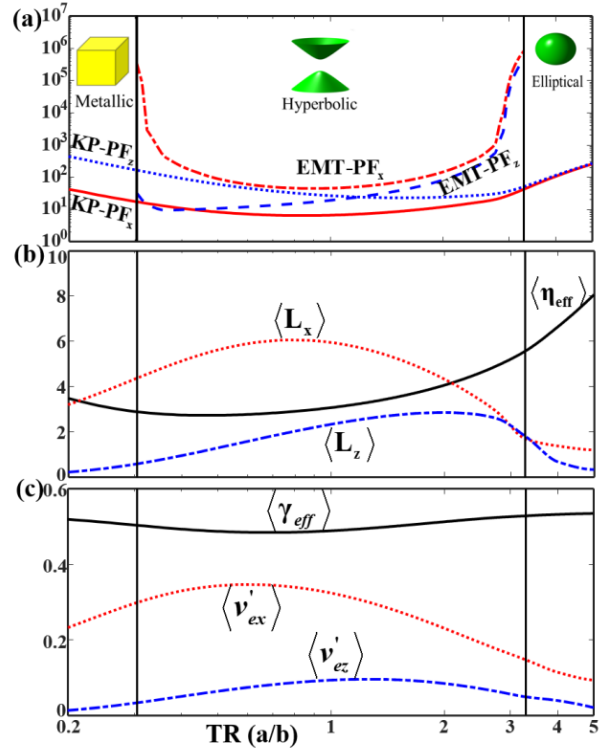

Fig. 5. (a) Comparison of EMT PF with the results of KP model; (b) Change of the mean loss, and energy velocity with *TR*; (c) Change of the mean impedance, and propagation length with *TR*.

The results are shown in Fig. 5. From Fig. 5(a) we can see that in the hyperbolic region PF estimated by the KP model is somewhat less than the one predicted by the EMT result. This is unsurprising and is simply the result of deviation of the KP IFS from the perfect hyperbola as shown in Fig. 2(b). What is more interesting is that strong PE not only exists outside of the hyperbolic region, but is substantially higher there than in the hyperbolic region. The PE appears to have very little to do with nature of dispersion. Apparently all the enhancement occurs "locally" and can be treated as enhancement by weakly coupled gap (metallic region) or slab (dielectric region) SPPs. Turning our attention to the mean parameters shown in Fig. 5(b) and (c) we once again see a familiar picture. As configurations yielding largest PE (of either very large or very small TR) also incur the highest loss, and have disappointingly high mean impedance and short propagation distances, they are always less than a micrometer in the lateral direction and far less than that in the more important for the extraction of radiation normal to the plane direction. It is clear that the energy simply couples into the large k-vector waveguide modes traveling along the plane and that these modes are no different from the short range modes in gap and slab SPPs. The emitter energy gets coupled into the kinetic motion of electrons in the metal and then dissipated. In essence what is observed is simply a quenching of radiative lifetime of the emitter [53].

## 6. THE IMPACT OF GRANULARITY

As mentioned in [31] the EMT does offer a good guidance for HMMs properties, yet fails to take their granularity into account. Fig. 6(a) shows the IFS for different periods when $TR = 1$, and the granularity is defined as $G = (a+b)/\lambda_d$. Here we can see that the IFS is strongly dependent on the granularity. For smaller granularity, the IFS get closer to EMT, but for large $k$ wavevector, the difference persists. Fig. 6(b) shows the relation between the maximum PF in the dielectric and the granularity for the same TR as in Fig. 6(a). As expected, extension of the BZ increase density of states and PF. But, as discussed above, the "new" large wavevector states are the ones with the large loss. Hence, while the radiative lifetime is expected to shorten even further with a decrease in period, the external efficiency would also decrease. In essence, shortening the period will only increase the quenching.

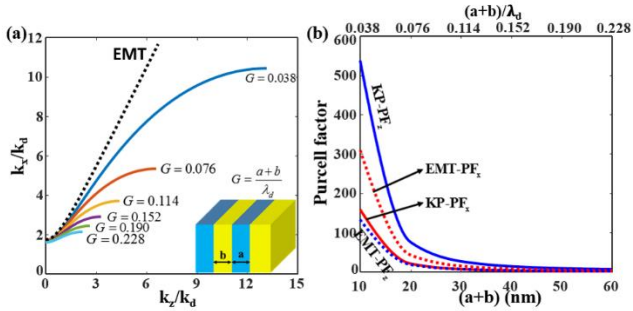

Fig. 6. (a) IFS for different periods; (b) Change of maximum PF for the two components of the emitting dipole in the dielectric with granularity.

## 7. COMPARISON OF SALB AND GAP SPPS

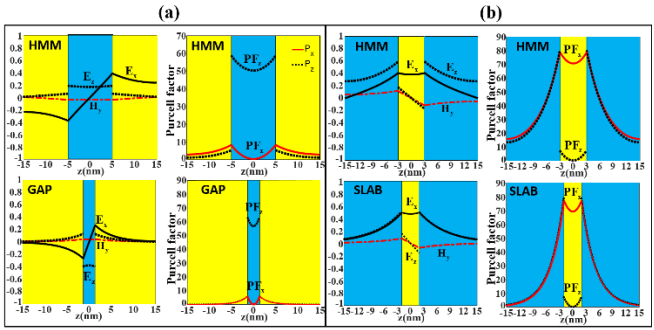

Fig. 7. Comparison of HMMs with dielectric gap waveguide (a) and metal slab waveguide (b).

To ascertain the relation between HMMs and SPPs, we have compared the HMMs with dielectric gap and metal slab plasmonic waveguides, referred here as the gap and slab SPPs respectively. Fig. 7(a) shows the comparison of the fields (left) and PF (right) of the HMMs (top) and gap SPPs waveguide (bottom). We can see that the fields are quite similar, with the normal electric field confined strongly inside the dielectric gap, leading to strong PE. The only difference is that, for gap SPPs waveguide, a smaller thickness is required to achieve the same PF with HMMs. The comparison between HMMs (top) and slab SPPs waveguide (bottom) in Fig. 7(b) follows the same storyline except the resemblance is even stronger. It is important to note that the magnetic field inside the metal slab changes sign, indicating that the mode is anti-symmetric or so-called short range SPPs [49]. Therefore, one can say that as $TR$ increases and layered material changes its character from metallic to hyperbolic to elliptical (dielectric) the nature of the "hyperbolic" or "large –k –vector" mode gradually changes from the symmetric mode of the gap SPPs to the asymmetric mode of the slab SPPs. One can also show that the elliptical (small k-vectors) mode corresponds to the long range or symmetric slab SPPs mode.

Next we evaluate the PF and effective parameters of the short range mode in slab SPPs as a function of metal thickness $d$ as shown in Fig. 8. As expected, as the thickness decreases the PF increases because the mode gets more confined and the group velocity decreases. It is easy to see that by reducing thickness one can always match and surpass the PF in HMMs. Just as in HMMs, this increase of PF is always accompanied by rapid decrease in propagation length and increase of effective impedance indicating that what appears to be a genuine enhancement of the emission is in reality just a quenching.

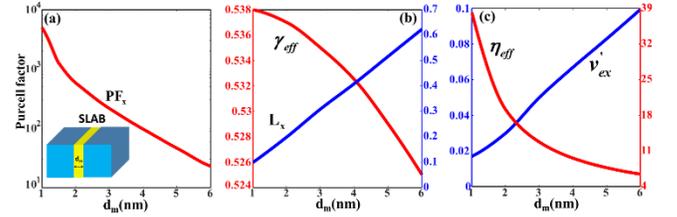

Fig. 8. Change of PF and effective parameters of metal slab with thickness of Ag, (a) Purcell factor; (b) Propagation length and effective loss; (c) Energy velocity and effective impedance.

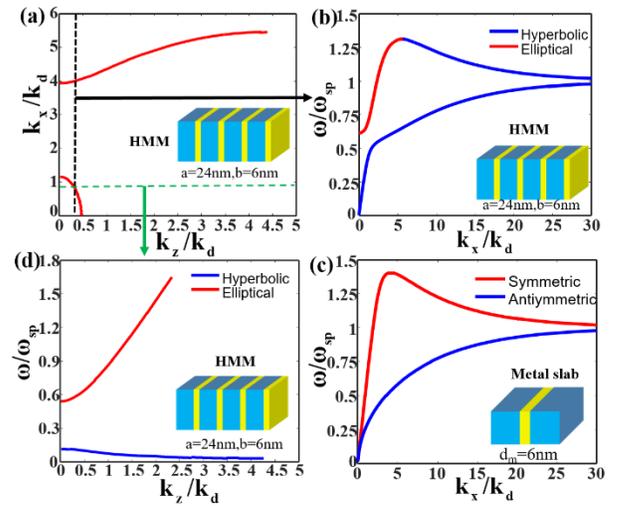

Fig. 9. (a) IFS at l = 500nm when the thickness of dielectric and metal are 24nm and 6nm respectively; (b)The lateral dispersion relation for HMMs in (a); (c)the dispersion relation of metal slab waveguide when the thickness of metal is 6nm; (d) The normal dispersion relation for HMMs in (a).

In order to further confirm the similarity between HMMs and SPPs waveguide, we have also calculated and compared their dispersion curves. In Fig. 9(a) the IFS at $\lambda = 500 nm$ for the 24 nm $Al_2O_3$ /6nm Ag layered structure is shown with both elliptic and hyperbolic branches present at this wavelength. By fixing the normal component of the wavevector at $k_z^{'} = 0.4$ (vertical dashed line in Fig. 9(a)), one can find the values of the lateral wavevector for the range of frequencies from 0 to $1.25\omega_{sp}$, where $\omega_{sp} = \omega_p / \sqrt{1+\varepsilon_d}$ is the surface plasma frequency, and produce the lateral dispersion curves of Fig. 9(b), that look remarkably similar to the dispersion curves of slab SPPs, shown in Fig. 9(c) below. The upper branch, which depending on wavevector, can correspond to elliptical or hyperbolic IFS obviously originates from the coupled long-range (symmetric) slab SPPs modes. The lower branch, always hyperbolic in nature corresponds to the short range (anti-symmetric mode). On the other hand, by fixing the lateral component of wave vector at $k_x^{'} = 0.8$ (horizontal dashed line in Fig. 9(a)), one obtains the dispersion curves in normal direction of Fig. 9(d) which looks precisely how one would expect to see weakly coupled modes of slab SPPs waveguides. Using a condensed matter analogy, in this tight binding approximation the elliptical curve looks like a "conduction band" and the hyperbolic curve as "valence band". The curvature of the hyperbolic band is much smaller than that of elliptical band indicating low velocity of propagation and larger PE.

## 8. BANDWIDTH OF PURCELL ENHANCEMENT

One of the purported advantages of HMMs can be considered the fact that the PE in them can be achieved over wide range of frequencies [20, 28, 29], unlike the enhancements near the single metal/semiconductor interface that are attainable only in the vicinity of the SP resonance. Indeed, as shown in Fig. 10 (solid curves), the range of PE is rather wide for HMMs with substantial enhancements over nearly an octave. However, the enhancement in the short range slab SPPs (two closely-spaced dashed curves) is also spread over a range that is almost as wide. Therefore, one can easily engineer the PF over the broad wavelength range by simply varying the thickness of slab SPPs (or width of gap SPPs) waveguide without resorting to fabrication of a multilayer structure. Once again, we stress the fact that most of the enhancement amounts to the quenching of radiative lifetime. Note also that in both HMMs and SPPs one can increase out-coupling of the high impedance modes with high k to some degree by using a grating or simply rough surface with virtually identical results [54].

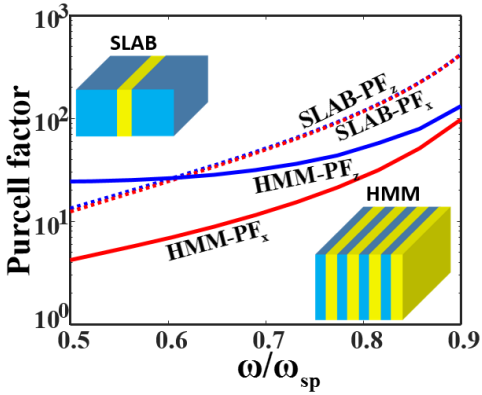

Fig. 10. Change of PF with frequency for HMMs (solid line) and metal slab SPPs waveguide (dashed line).

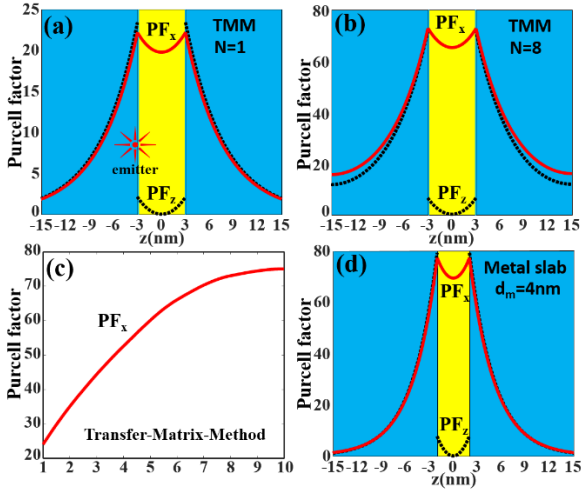

Fig. 11. (a-b) PF calculated using transfer-matrix-method for different periods; (d) Change of PF with the increase of period; (d) The PF obtained by metal slab with smaller thickness can be the same as HMMs in Fig. 7(b) (top).

It is also interesting to see how many alternating metal/dielectric layers are required to fully achieve HMMs behavior. We have performed the analysis (see Section 6 of Supplement 1) using Transfer Matrix method and have found out that after about 8 periods the characteristics of the structure no longer change, as shown in Fig. 11. But, as shown above, the same characteristics can also be obtained with just alternating layers in either gap SPPs or short range slab SPPs configuration.

## 9. THE PHYSICAL ORIGIN OF PURCELL ENHANCEMENT

Extensive discussion of characteristics of HMMs has led us to an unavoidable conclusion: the large wavevectors and, consequently, large density of states in these materials are all accompanied by large loss in the metal. In that respect, HMMs are no different from simple slab and gap SPPs, and, furthermore, the densities of states only get larger in the regions where the metamaterial is nominally "metallic" or "elliptical", i.e., they have very little to do with the "hyperbolicity". This conclusion is, of course, only logical as long as one accepts the obvious fact that the quantum "states" do not appear out of nowhere and their density can only be altered by coupling between the different states. In this respect, the density of photons always remains the same and it only changes in the dielectric medium because the photons couple with the polarization oscillations of atoms or molecules forming polaritons. This essentially add the "degrees of freedom". In the plasmonic structures, including HMMs, the photons couple with the collective oscillations of free carriers in the metal. The density of states for free electrons near the Fermi energy is roughly 8 orders of magnitude higher than density of photons, mostly because the electron velocity is much slower than the speed of light. As a result, new coupled plasmon-polariton modes have much larger density of states. In other words, the giant PF in any plasmonic structure including HMMs is simply the consequence of having large density of free carriers. To check this conjecture, we first note that the presence of a large number of moving free carriers is manifested by the large "normalized group dielectric constant" $\varepsilon_g' = \varepsilon_d^{-1} \partial(\omega \varepsilon_m)/\partial \omega \approx \varepsilon_d^{-1} \omega_p^2 / \omega^2$. It is this large derivative that causes plasmon polaritons to propagate slowly, which in turn, leads to large density of state and PF. What if one considers a hypothetical material with negative permittivity equal to that of actual metal ($\varepsilon_m = -12$) but dispersionless in the region of interest (such material of course cannot exist because the electric field energy in it would be negative). We plot two IFS separated by small frequency interval $\delta\omega \approx 0.9\% \omega_{sp}$ of this HMMs made up of "dispersionless metal" in Fig. 12 (a), next to the IFS of the HMMs made up with real Ag, whose permittivity has dispersion. The difference is dramatic – for real Ag the two curves diverge at large wavevectors, while for the dispersionless metal the IFS converges. This is expected even in the EM theory where without dispersion hyperbolic IFS of all frequencies converge to the same asymptote. As a result, the density of states without dispersion decreases dramatically, and, as shown in Fig. 12(b), so does the PF. One can see that about 90% of the giant PE comes simply from coupling of the emitter's energy into the kinetic motion of free electrons and owes preciously little to exactly how the layers are arranged. Needless to say, once the energy is coupled into the collective motion of free carriers it dissipates at the femtosecond rate and thus giant PF usually indicates a quenching of radiative lifetime. This discussion of the origin of giant PE in layered HMMs is also relevant to the natural HMs, such as hexagonal BN [55], where the modes with large wave vectors are of course nothing but phonon polaritons in which the energy is contained mostly not in the form of electromagnetic field but in the form of ionic vibrations (optical phonons). In other words, the energy of the emitter placed inside natural HMs is coupled directly into ionic vibrations. The ionic vibrations are of course damped, albeit not as strongly as free electrons (picoseconds vs. tens of femtoseconds), but then the density of states of these vibrations is also less than density of free electrons at Fermi level. Hence the basic trade-off between the large density of states and low loss is maintained in this material as well.

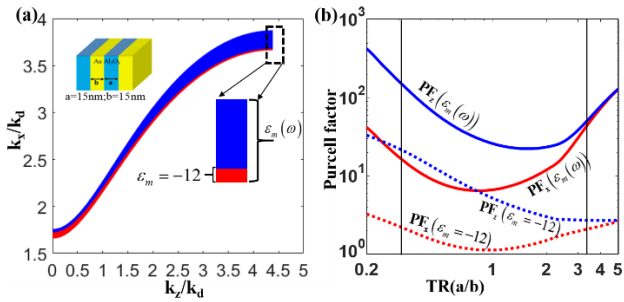

Fig. 12. Comparison of (a) IFSs and (b) PFs of the HMM with real metal and hypothetical dispersionless metal indicates that most of density of states and Purcell enhancement originates from the metal dispersion.

## 10. CONCLUSIONS

In this work we have investigated the optical properties of subwavelength layered metal/dielectric structures, also known as hyperbolic metamaterials, using exact and straightforward, fully analytical KP model. We have revealed a number of important features that have not been previously given proper attention. First of all, not only, as previously noted in [31] can hyperbolic and elliptical IFS co-exist, but hyperbolic IFS can exist for all combinations of layer permittivities and thicknesses. Most importantly, the largest PE of spontaneous radiation is achieved away from the hyperbolic region. Secondly, detailed comparison of the field distributions, dispersion curves, losses and PF between the HMMs and SPPs guided modes in metal/dielectric waveguides demonstrates that HMMs are nothing but the weakly coupled gap or slab SPPs modes. Large wave vectors and broadband PE are not specific to the HMMs and can be easily attained in thin metallic layers by varying their thickness. Third, we demonstrate that large wavevectors and PE in layered plasmonic structures are inextricably tied to the loss, slow group velocity, small propagation distances and large impedances. This indicates that the much heralded PE in the HMMs is actually direct coupling of the energy into the free electron motion in the metal, commonly known as quenching of radiative lifetime. There are far easier and well proven ways to modify the luminescence time such as adding defects and using low temperature grown materials [56]. Finally, looking deeper into the physics of PE in HMMs shows that it has very little to do with the hyperbolicity per se and everything to do with the large dispersion of permittivity in the metals or polar dielectrics, as our conclusions are relevant also for the naturally infrared HMs occurring in nature. It is our opinion, that while HMMs do present a fascinating research subject, at least when it comes to enhancement of radiating processes and field concentrations, HMMs have no significant advantages when compared to far simpler plasmonic structures.

**FUNDING INFORMATION**

National Science Foundation (NSF) (1507749); and Army Research Office (ARO) (W911NF-15-1-0629).

**ACKNOWLEDGMENTS**

We thank Dr. P. Noir for his insight during our scientific discussion and numerous suggestions.

See Supplement 1 for supporting content.